\documentclass[twocolumn]{aastex631}

\usepackage{hyperref}

\providecommand{\dodoi}[1]{doi:~\href{http://doi.org/#1}{\nolinkurl{#1}}}

\shorttitle{EM Characterization of ZTF J0526$+$5934}
\shortauthors{Kosakowski, Kupfer, Bergeron, \& Littenberg}
\graphicspath{{./}{figures/}}

\begin{document}

\title{Electromagnetic characterization of the LISA verification binary ZTF J0526$+$5934}

\correspondingauthor{Alekzander Kosakowski}
\email{a.kosakowski@protonmail.com}

\author[0000-0002-9878-1647]{Alekzander Kosakowski}
\affiliation{Department of Physics and Astronomy, Texas Tech University, 
2500 Broadway 
Lubbock, Texas 79409, USA}

\author[0000-0002-6540-1484]{Thomas Kupfer}
\affiliation{Department of Physics and Astronomy, Texas Tech University, 
2500 Broadway 
Lubbock, Texas 79409, USA}

\author[0000-0003-2368-345X]{P. Bergeron}
\affiliation{Departement de Physique, Universit\'{e} de Montr\'{e}al, C.P. 6128, Succ. Centre-Ville, Montr\'{e}al, Quebec H3C 3J7, Canada}

\author[0000-0002-9574-578X]{Tyson B. Littenberg}
\affiliation{NASA Marshall Space Flight Center, Huntsville, Alabama 35811, USA}

\begin{abstract}
We present an analysis of new and archival data to the 20.506-minute LISA verification binary J052610.42$+$593445.32 (J0526$+$5934). Our joint spectroscopic and photometric analysis finds that the binary contains an unseen $M_1=0.89\pm0.11~{\rm M_\odot}$ CO-core white dwarf primary with an $M_2=0.38\pm0.07~{\rm M_\odot}$ post-core-burning subdwarf, or low-mass white dwarf, companion. Given the short orbital period and relatively large total binary mass, we find that LISA will detect this binary with signal-to-noise ratio $44$ after 4 years of observations. J0526$+$5934 is expected to merge within $1.8\pm0.3~{\rm Myr}$ and likely result in a ${\rm D}^6$ scenario Type Ia supernova or form a He-rich star which will evolve into a massive single white dwarf.
\end{abstract}


\section{Introduction}

White dwarfs represent a relatively simple final evolutionary stage for most single-star stellar evolution. Interactions in a binary system complicate this evolution and can result in a wide range of astrophysically interesting systems. For binary evolution, the more massive star will evolve first, potentially leading to a phase common-envelope evolution as it evolves onto the asymptotic giant branch. This process strips the primary of its outer layers and leaves behind a CO-core white dwarf in a compact binary with orbital period ranging from hours to days. Depending on the mass ratio of the resulting compact binary, a second common-envelope phase may occur as the companion fills its Roche lobe near the base of the red giant branch. This double-common-envelope evolutionary process results in a double-degenerate binary with orbital period ranging from less than an hour to a only a few hours \citep{li2019}. Compact post-common-envelope binaries are excellent systems for studying binary evolution. Recent work by \citet{scherbak2023} used compact eclipsing white dwarf binaries to place constraints on the common envelope ejection efficiency.

Compact binaries with periods less than about $6~{\rm h}$ are considered to be merging binaries since the rate of their orbital angular momentum loss caused by gravitational wave emission is sufficient to result in a binary merger within a Hubble time. The merging binaries observed today therefore represent a population of progenitor binaries to merger products, such as AM CVn binaries \citep{kilic2016}, He-rich stars \citep{zhang2014}, massive single white dwarfs \citep{cheng2020, kilic2023}, and Type Ia supernovae \citep{woosley1986, fink2007, liu2018, shen2018a}. Characterization of merging white dwarf binaries provides constraints on the formation rates and potential formation channels of these merger products. Many compact white dwarf binaries have been discovered through targeted spectroscopic surveys, such as the ELM Survey \citep{brown2010,brown2022,kosakowski2023}, and through large-scale systematic searches for photometric variability in time-domain surveys \citep{burdge2020,vanroestel2022,ren2023}.

\begin{figure*}
    \centering
    \includegraphics[scale=0.483]{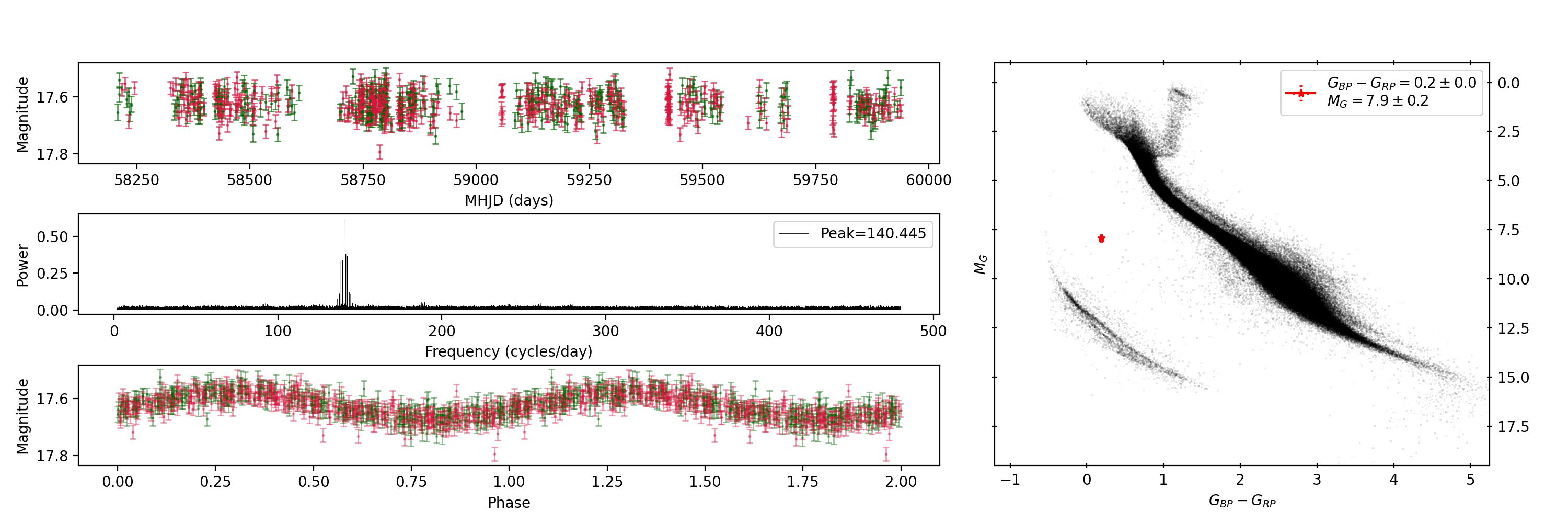}
    \caption{Left: ZTF DR16 light curve of J0526$+$5934 (top), its Lomb-Scargle power spectrum (middle), and phase-folded ZTF DR16 light curve (bottom). Data points are colored based on the filter used. Green data points represent ZTF\_$g$, red data points represent ZTF\_$r$. Right: Gaia DR3 color-magnitude diagram. The location of J0526$+$5934 is marked with a red symbol.}
    \label{fig:0526_ztf}
\end{figure*}

White dwarf binaries are expected to be the dominant source of gravitational wave signal for the Laser Interferometer Space Antenna \citep[LISA;][]{amaro2017}. The shortest period binaries, with $P\lesssim1~{\rm h}$, emit gravitational waves at mHz frequencies that may be detected by LISA. LISA is expected to detect $\mathcal{O}(10^4)$ of these ultra-compact binaries, but only $\mathcal{O}(10^2)$ are expected to also be detectable through their electromagnetic radiation, allowing for a multi-messenger approach to studying binary evolution \citep{nelemans2001,korol2017,li2020,amaro2023}. The strongest gravitational wave emitters will act as ``verification binaries," which can be used to calibrate the LISA data set in the first few months of operation. So far, about 40 LISA detectable binaries have been characterized through their electromagnetic radiation \citep[see][and references therein]{finch2023,kupfer2023}.

Here we present an independent discovery and analysis of a new LISA verification binary with orbital period $P=20.506~{\rm min}$, J052610.42+593445.32 (J0526$+$5934). J0526$+$5934 was originally reported as a candidate ultra-compact binary by \citet{ren2023} based on periodic photometric variability seen in the Zwicky Transient Facility \citep[ZTF;][]{bellm2019,graham2019,masci2019} data archive. The authors find that J0526$+$5934 will be detected by LISA with an expected signal-to-noise ratio ${\rm S/N}=35.788$ after 4 years of observation.

Throughout this work, we adopt the convention that the unseen massive star, which evolved first, is the primary star, while the relatively low mass companion is the secondary, such that $M_1>M_2$. In Section \ref{sec:targetselect} we describe our target selection criteria. In Sections \ref{sec:spec} and \ref{sec:phot} we provide the details of our spectroscopic and photometric analysis. In Sections \ref{sec:decay} and \ref{sec:discussion} we discuss the expected rate of orbital decay of J0526$+$5934, prospects for LISA detection, and its potential merger outcomes. 
Finally, we summarize our results in Section \ref{sec:conc}.

\section{Target selection}\label{sec:targetselect}

We selected all targets from the Gaia eDR3 \citep{gaia_edr3} white dwarf catalog \citep{gentile2021} and performed a generalized period search on their associated ZTF DR10 archival light curves using the \textsc{astropy} \citep{astropy2022} implementation of the Lomb-Scargle periodogram \citep{lomb1976,scargle1982,vanderplas2018}. We searched for periodic signals with periods between $P_{\rm min}=3~{\rm min}$ and $P_{\rm max}=684~{\rm min}$, split into 10-million evenly-spaced trial frequencies. To increase temporal sampling of the ZTF light curves with multiple measurements in different filters, we median-combined the light curves across each filter by artificially shifting the $r$- and $i$-band data such that their median magnitude values matched the median $g$-band magnitude. Our search made use of the Texas Tech University High Performance Computing Center to efficiently process each light curve. We manually inspected the output light curve images to identify objects with periodic photometric variability based on their peak power spectrum value with respect to the local noise level.

J0526$+$5934 (Gaia DR3 282679289838317184) was identified in our search as an ultra-compact binary, with dominant frequency $f_{\rm peak}\approx140.445~{\rm cycles~d^{-1}}$ ($P_{\rm peak}\approx10.253~{\rm min}$) and amplitude $A\approx~0.05~{\rm mag}$, suggesting ellipsoidal modulation at true orbital period $P\approx20.506~{\rm min}$. We estimated the uncertainty in the orbital period through a bootstrapped analysis with 10,000 periodograms of the ZTF DR16 light curve data focused on the surrounding 40-seconds of the most-probable period, split into 20-million frequency bins, and find $P=1230.37467\pm0.00007~{\rm s}$. Figure \ref{fig:0526_ztf} presents the ZTF light curve (left:top), its Lomb-Scargle power spectrum (left:middle), and the ZTF light curve phase-folded at the most-probable frequency (left:bottom). We mark the location of J0526$+$5934 on the Gaia DR3 \citep{gaia_dr3} color-magnitude diagram (right) as a red star.

Our generalized search on the ZTF DR10 archival data recovered a few previously published ultra-compact binaries, including J0651$+$2844 \citep[$P=12.8~{\rm min}$;][]{brown2011}, ZTF J0538$+$1953 \citep[$P=14.4~{\rm min}$;][]{burdge2020}, PTF J0533$+$0209 \citep[$P=20.6~{\rm min}$;][]{burdge2019b}, ZTF J0722$-$1839 \citep[$P=23.7~{\rm min}$;][]{burdge2020}, ZTF J1901$+$5309 \citep[$P=40.6~{\rm min}$;][]{burdge2020}, J2049$+$3351 \citep[$P=42.8~{\rm min}$;][]{kosakowski2023}, and ZTF J2320$+$3750 \citep[$P=55.3~{\rm min}$;][]{burdge2020}. Our algorithm failing to recover other well-characterized binaries may be a consequence of our evenly-spaced frequency grid, which may be under-sampled at higher frequencies and over-sampled at lower frequencies. However, some of these binaries, such as ZTF J1539$+$5027 \citep[$P=6.9~{\rm min}$;][]{burdge2019a}, are not included in the Gaia eDR3 white dwarf catalog and therefore are not identified in our search, but are otherwise easily recovered with our algorithm when targeted.

\section{Spectroscopic Analysis}\label{sec:spec}
\subsection{Keck Archival Spectroscopy}
J0526$+$5934 was originally observed on UT 2020 September 16 with the Keck 10-meter telescope on Maunakea as part of the program ID 2020B-C282\footnote{Dataset \dodoi{10.26135/KOA2}} (PI: Prince). The observations used LRIS \citep{oke1995} with the blue-channel 600/4000 grism ($600~{\rm lines~mm^{-1}}$; $\lambda_0=4000~{\rm \AA}$), 1.0\arcsec\ slit, and $2\times2$ CCD binning, providing a spectral resolution of $\approx4.0~{\rm \AA}$ over the wavelength range $3250\sim5550~{\rm \AA}$. These observations include ten consecutive spectra with 120-second exposures over approximately one full binary orbit.

We downloaded the blue optical spectra and their associated calibration data from the Keck Observatory Archive and reduced the data using using standard \textsc{iraf} \citep{tody1993} procedures including image correction, spectral extraction, dispersion correction using HgNeArCdZn arc-lamps, and wavelength calibration using BD28$^\circ$4211 standard star observations taken with the same setup.

\begin{table}
\center
  \renewcommand{\arraystretch}{1.15} 
  \addtolength{\tabcolsep}{2pt}
	\begin{tabular}{l c}
    \hline
    \hline
    \multicolumn{1}{C}{\rm MJD} &
    \multicolumn{1}{C}{$v_{\rm r}$} \\
    \multicolumn{1}{C}{(${\rm d}$)} &
    \multicolumn{1}{C}{(${\rm km~s^{-1}}$)} \\
    \hline
59108.569310 & $ 497.88\pm2.99   $ \\
59108.571125 & $ 289.56\pm9.29   $ \\
59108.572941 & $-131.02\pm7.99   $ \\
59108.574759 & $-494.01\pm11.18  $ \\
59108.576571 & $-593.67\pm7.84   $ \\
59108.578389 & $-331.64\pm6.56   $ \\
59108.580201 & $  94.91\pm8.08   $ \\
59108.582019 & $ 442.35\pm15.86  $ \\
59108.583831 & $ 504.10\pm6.96   $ \\
59108.585649 & $ 233.46\pm10.03  $ \\
    \hline
    \hline
	\end{tabular}
    \caption{Radial velocity measurements for J0526$+$5934 based on our cross-correlation fit.}
    \label{table:rv_table}
\end{table}
The optical spectrum of J0526$+$5934 is dominated by hydrogen absorption features and has relatively shallow He I absorption features at $4912~{\rm \AA}$, $4471~{\rm \AA}$, and $4026~{\rm \AA}$, giving it the DAB classification. We see no evidence of the companion in the Keck spectroscopy.

\newpage
\subsection{Radial Velocity and Kinematics}
We estimated the radial velocity for each of the ten blue optical spectra against a zero-velocity low-mass DA white dwarf template spectrum using the cross-correlation package \textsc{rvsao.xcsao} \citep{kurtz1998} within \textsc{iraf}. We then shifted each of the ten component spectra of J0526$+$5934 to zero-velocity and co-added them into a single high-quality, zero-velocity spectrum, which we later use to estimate atmospheric parameters. Finally, we obtained precise radial velocity estimates for each component spectrum by using the co-added spectrum as a template for another round of cross-correlation. Our individual radial velocity measurements are presented in Table \ref{table:rv_table}.
\begin{figure}
    \centering
    \includegraphics[scale=0.4]{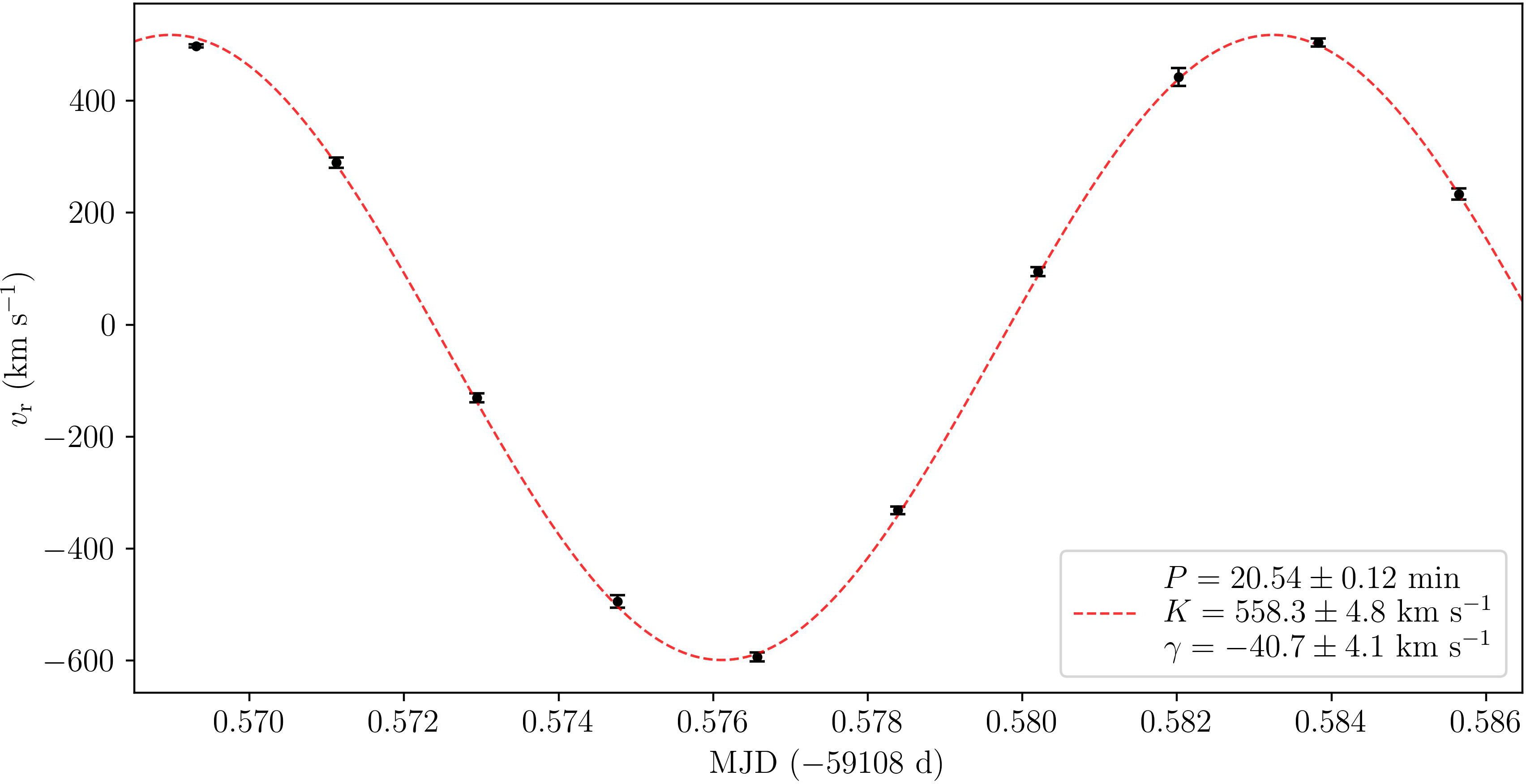}
    \caption{Best-fitting exposure-averaged integrated sine curve (red dashed line; ${\rm EXPTIME=120}~{\rm s}$) to the Keck LRIS radial velocity measurements of ZTF J0526$+$5934 (black points).}
    \label{fig:keck_rv}
\end{figure}

We fit a circular orbit to the radial velocity measurements to estimate the orbital period ($P$), velocity semi-amplitude ($K$), and systemic velocity ($\gamma$) of the binary. We find best-fitting parameters $P_{\rm RV}=20.54\pm0.12~{\rm min}$, $K=549.7\pm4.7~{\rm km~s^{-1}}$, and $\gamma=-40.7\pm4.1~{\rm km~s^{-1}}$, roughly twice the orbital period identified through our Lomb-Scargle analysis of the ZTF light curve. However, because the exposure time used for each spectrum covers a significant fraction of the orbital period ($9.8\%$), we corrected the orbital solution by fitting an average integrated sine curve to the observed data, taking into account the exposure time at each observed orbital phase. We find smearing-corrected velocity semi-amplitude $K=558.3\pm4.8~{\rm km~s^{-1}}$, corresponding to mass function $0.255\pm0.007~{\rm M_\odot}$. Our best-fitting average integrated sine curve orbital solution is presented in Figure \ref{fig:keck_rv}.

We estimated Galactic space velocities for J0526$+$5934 by using our best-fitting systemic velocity and the Gaia DR3 astrometry measurements. We find $U=47.6\pm1.9~{\rm km~s^{-1}}$ ($U$ positive toward the Galactic center), $V=-7.3\pm1.7~{\rm km~s^{-1}}$, and $W=3.8\pm1.1~{\rm km~s^{-1}}$, corrected for the motion of the local standard of rest \citep{schonrich2010}, consistent with a Galactic disk population based on the average velocity and dispersion distributions for the Galactic disk and halo from \citet{chiba2000}.

\subsection{Atmospheric Parameters}
\begin{figure}
    \centering
    \includegraphics[scale=0.35]{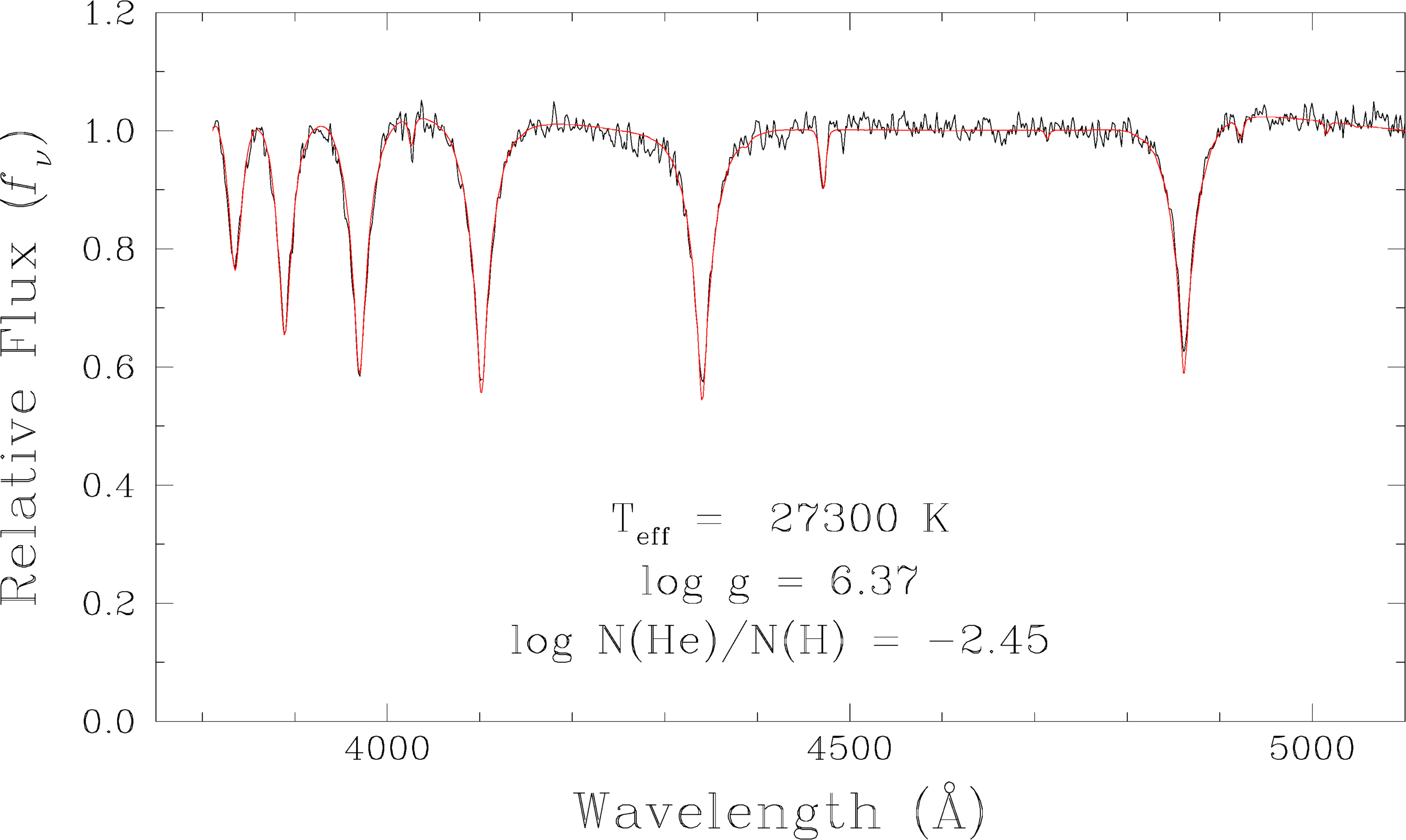}
    \caption{Best-fitting model atmosphere to the co-added optical spectrum of ZTF J0526$+$5934.}
    \label{fig:keck_he_fit}
\end{figure}

We estimated the atmospheric parameters of J0526$+$5934 by fitting a grid of hot subdwarf model atmospheres \citep{saffer1994} to the co-added blue optical spectrum and find best-fitting parameters $T_{\rm eff}=27300\pm260~{\rm K}$, $\log{g}=6.37\pm0.03$, and $\log{\frac{N(He)}{N(H)}}=-2.45\pm0.06$, which suggest that J0526$+$5934 is a post-core-burning subdwarf, or an inflated He-core low-mass white dwarf. We summarize our best-fitting parameters in Table \ref{table:main_table}. Our best-fitting model is over-plotted onto the Keck blue optical spectrum in Figure \ref{fig:keck_he_fit}.

\newpage
\section{Photometric Analysis}\label{sec:phot}
\subsection{Spectral Energy Distribution}
A spectral energy distribution (SED) fit was performed to measure the radius and mass of J0526$+$5934. The angular diameter of the star is measured and, in combination with the Gaia DR3 parallax, we derive the radius of the visible component in J0526$+$5934. The luminosity and mass are calculated using the atmospheric parameters measured from spectroscopy. This method is described in detail by \citet{Heber2018}. Because J0526$+$5934 is missing archival GALEX UV and SDSS $u$-band photometry, we fixed the effective temperature and surface gravity to our spectroscopic values.

Using the functions of \citet{Fitzpatrick2019}, we account for interstellar reddening. The color excess $E$\,(44$-$55) is treated as a free parameter and the the extinction parameter $R(55)$ was fixed to the standard value of 3.02. To estimate the radius we apply $R=\Theta/2\varpi$, where $\Theta$ is the angular diameter derived from the SED fit and $\varpi$ is the parallax extracted from Gaia DR3. The mass follows from the $M=g R^2/G$, where $g$ is the surface gravity and $G$ is the gravitational constant. Our fit to the available SED, including Gaia $G$, $G_{\rm BP}$, and $G_{\rm RP}$, PanSTARRS $grizy$ \citep{chambers2016}, and (un)WISE $W1$ \citep{schlafly2019}, finds $R_2=0.061^{+0.006}_{-0.005}~{\rm R_\odot}$, corresponding to mass $M_2=0.32^{+0.06}_{-0.05}~{\rm M_\odot}$.

\begin{figure*}
    \centering
    \includegraphics[scale=0.25]{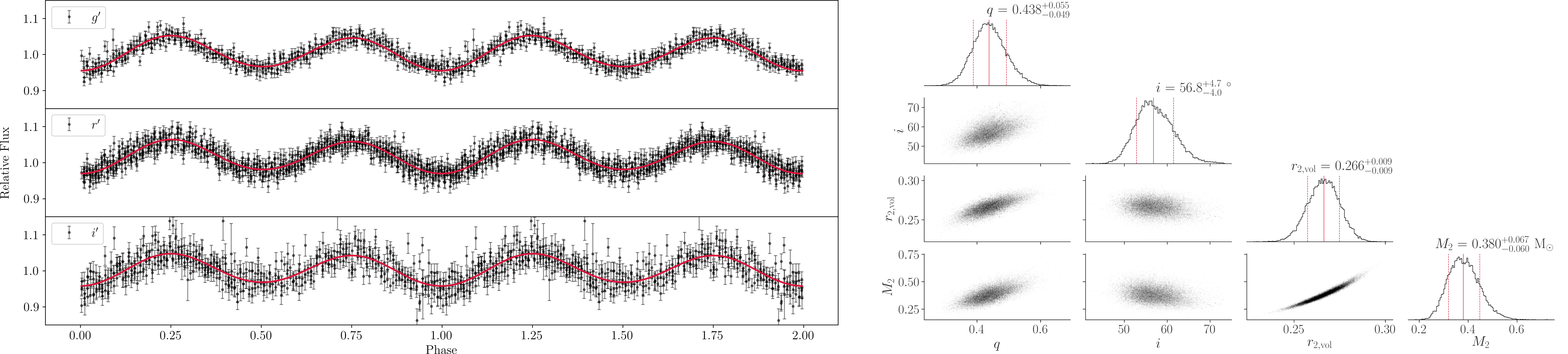}
    \caption{Left: Best-fitting \textsc{lcurve} models over-plotted onto the phase-folded McDonald 2.1-meter telescope $g'$-band (top), $r'$-band (middle), and $i'$-band (bottom) light curves. Right: Parameter distributions for the mass ratio ($q=M_2/M_1<1.0$), orbital inclination ($i$), volumetric scaled stellar radius ($r_{\rm 2,{\rm vol}}=R_{2,{\rm vol}}/a$), and stellar mass $M_2$. We mark the upper and lower $1\sigma$ error values as the $84.13$ and $15.97$ percentiles to each parameter distribution, respectively.}
    \label{fig:combined_lcurve}
\end{figure*}
\begin{table}
\center
  \renewcommand{\arraystretch}{1.5}
  \addtolength{\tabcolsep}{2pt}
	\begin{tabular}{l r}
    \hline
    \hline
    Source ID (Gaia DR3) & 282679289838317184 \\
    R.A. (2016.0) & 05:26:10.420 \\
    Decl. (2016.0) & +59:34:45.318 \\
    Gaia G (mag) & $17.563\pm0.003$ \\
    Parallax (mas) & $1.18\pm0.09$ \\
    \hline
    $P_{\rm ZTF}~{\rm (s)}$ & $1230.37467\pm0.00007$ \\
    $\dot{P}_{\rm expected}~{\rm (s~s^{-1})}$ & $-(8.66\pm3.03)\times10^{-12}$ \\
    \hline
    {$T_{\rm eff}~{\rm (K)}$} & {$27,300\pm260$} \\
    {$\log{g}~{\rm (cm~s^{-2})}$} & {$6.37\pm0.03$} \\
    {$\log{\frac{N(He)}{N(H)}}$} & {$-2.45\pm0.06$} \\
    {$K~{\rm (km~s^{-1})}$} & {$558.3\pm4.8$} \\
    {$\gamma~{\rm (km~s^{-1})}$} & {$-40.7\pm4.1$} \\
    {Mass Function (${\rm M_\odot}$)} & $0.255\pm0.007$ \\
    \hline
    {$q=\frac{M_2}{M_1}$} & {$0.426^{+0.052}_{-0.051}$} \\
    {$r_{2}=\frac{R_2}{a}$} & {$0.288\pm0.011$} \\
    {$r_{2,\rm vol}$} & {$0.264\pm0.009$} \\
    {$i~{\rm (^\circ)}$} & {$57.1^{+4.3}_{-4.1}$} \\
    {$M_2~{\rm (M_\odot)}$} & {$0.378^{+0.066}_{-0.060}$} \\
    {$M_1~{\rm (M_\odot)}$} & {$0.887^{+0.110}_{-0.098}$} \\
    {$a~{\rm (R_\odot)}$} & {$0.266^{+0.011}_{-0.010}$} \\
    {$R_{2,\rm vol}~{\rm (R_\odot)}$} & {$0.070\pm0.005$} \\
    {$T_0~{\rm (d^{+s}_{-s})}$} & {$2459854.910239^{+1.2}_{-1.3}$} \\
    \hline
    \hline
	\end{tabular}
    \caption{Fitted and archival parameter values for J0526$+$5934.}
    \label{table:main_table}
\end{table}
\subsection{Light Curve Modeling}
We obtained high-speed $g'$, $r'$, and $i'$-band follow-up light curves of J0526$+$5934 using the McDonald 2.1-meter telescope on 2022 September 30, 2022 October 01, and 2022 October 02, respectively.

We used \textsc{lcurve} \citep{copperwheat2010} to perform simultaneous $g'$-,  $r'$-, and $i'$-band modeling to our McDonald light curves. We fit for the mass ratio ($q=M_2/M_1<1.0$), orbital inclination ($i$), scaled companion radius ($r_2=R_2/a$), time of primary conjunction ($t_0$), and the filter-dependent gravity-darkening and quadratic limb-darkening coefficients. We included Gaussian priors on the surface gravity, effective temperature, velocity semi-amplitude, and radius of the low-mass companion based on the values obtained from our fits to the optical spectroscopy and available SED. We used gravity and quadratic limb-darkening coefficients from \citet{claret2020} for DA white dwarfs with atmospheric parameters $T_{\rm eff,2}=27,500~{\rm K}$, $\log{g_2}=6.37$ and $T_{\rm eff,1}=10,000~{\rm K}$, $\log{g_1}=8.00$. We marginalized over the limb and gravity-darkening coefficients by assigning Gaussian priors based on the $2\sigma$ uncertainties of our spectroscopic atmospheric parameters.

We find most-probable model parameters $q=0.426^{+0.052}_{-0.051}$, $i=57.1^{+4.3}_{-4.1}$$^\circ$, and $R_{\rm 2,vol}=0.070\pm0.005~{\rm R_\odot}$, where $R_{\rm vol}$ is the volumetric radius. These parameters correspond to stellar masses $M_2=0.378^{+0.066}_{-0.060}~{\rm M_\odot}$ and $M_1=0.887^{+0.110}_{-0.098}~{\rm M_\odot}$, in agreement to within $1\sigma$ of the mass and radius estimates from our SED fitting. We adopt the light curve modeling solution as the true mass and radius and summarize these parameters in Table \ref{table:main_table}. Figure \ref{fig:combined_lcurve} presents a corner-plot of our parameter distributions with the the most-probable model over-plotted onto our McDonald light curves.

\section{Orbital Decay}\label{sec:decay}
The orbit of compact binaries decays due to the loss of orbital angular momentum through the emission of gravitational waves. We estimated the magnitude of this effect for J0526$+$5934 using Equation \ref{eqn:pdot2} \citep{landau1975,piro2019}

\begin{equation}
\dot{P}_{\rm GW}=-\frac{96}{5}\frac{G^{3}}{c^5}\frac{M_1M_2(M_1+M_2)}{a^4}P_{\rm orb}
\label{eqn:pdot2}
\end{equation}
where $a$ is the binary separation. We find that the expected rate of orbital decay in J0526$+$5934 due to the emission of gravitational waves is $\dot{P}_{\rm GW}=-(8.25\pm2.88)\times10^{-12}~{\rm s~s^{-1}}$, where the large uncertainties are dominated by our uncertainty in the component masses.

Tidal interactions contribute to the total orbital decay in ultra-compact binaries as orbital energy is used to spin-up the stars in the binary. We ignore the effects of tidal heating and assumed that the stars are tidally locked and estimated the contribution from tidal interactions to the orbital decay of J0526$+$5934 using Equation \ref{eqn:pdot_tide} \citep[see Equation 6 in][]{piro2019}
\begin{equation}
\dot{P}_{\rm total}=\dot{P}_{\rm GW}\left[1-3\frac{(I_1+I_2)}{a^2}\frac{(M_1+M_2)}{M_1M_2}\right]^{-1}
\label{eqn:pdot_tide}
\end{equation}
where $I_i=k_iM_iR_i^2$ is the moment of inertia of each star. \citet{burdge2019b} finds $k_2=0.066$ and $k_1=0.14$ based on white dwarf models for less massive stellar components in an ultra-compact binary, while \citet{marsh2004} finds that $k\approx0.2$ is an appropriate estimate for white dwarfs based on the Eggleton zero-temperature mass-radius relation. We used $k_1=k_2=0.15\pm0.05$ and find total orbital decay $\dot{P}_{\rm GW}+\dot{P}_{\rm tides}=-(8.66\pm3.03)\times10^{-12}~{\rm s~s^{-1}}$, corresponding to tidal contribution $\frac{\dot{P}_{\rm tides}}{\dot{P}_{\rm total}}=4.8\pm1.9\%$.

The orbital decay of compact binaries can be directly measured as an observable through timing offsets in periodic photometric variability, such as with precise eclipse timing measurements over multi-year baselines \citep[see][]{hermes2012,burdge2019a,burdge2019b,burdge2020,burdge2023}. However, while we find a baseline of $\approx2500~{\rm d}$ in archival data from ZTF and the Asteroid Terrestrial-impact Last Alert System \citep[ATLAS;][]{tonry2018,heinze2018}, the orbital period precision in our McDonald data is insufficient to measure the effects of orbital decay in J0526$+$5934. Future long-term monitoring of J0526$+$5934 will provide precise orbital period and ephemeris timings which will be used to directly measure the orbital decay and place observational constraints on the chirp mass of J0526$+$5934.

\section{Discussion}\label{sec:discussion}
\subsection{LISA Detection}
We used the parallel tempered Markov Chain Monte Carlo algorithm \textsc{gbmcmc} within \textsc{ldasoft} \citep{littenberg2020} to simulate the expected gravitational wave signal of J0526$+$5924 from LISA. We fixed the sky position and orbital period, placed a Gaussian prior on the distance based on the Gaia DR3 parallax, and placed a uniform prior on the orbital inclination based on the $2\sigma$ uncertainties from our light curve modeling. Our simulations find that LISA will recover the orbital inclination and chirp mass with similar or better precision than our electromagnetic analysis after a 2-year mission with inclination $i=56.3^{+3.7}_{-5.2}$$^\circ$, chirp mass $\mathcal{M}=0.49\pm0.03~{\rm M_\odot}$,
and gravitational wave amplitude $\mathcal{A}=(2.9^{+0.3}_{-0.2})\times10^{-22}$, corresponding to signal to noise ratio ${\rm S/N}\approx27$ after a 2-year mission, and ${\rm S/N}\approx44$ after a 4-year mission, using the Galactic foreground noise model of \citet{cornish2017}.

\subsection{Merger Outcome}
We find that J0526$+$5934 will merge within $\tau_{\rm GW}=1.8\pm0.3~{\rm Myr}$ due to loss of orbital angular momentum from a combination of gravitational wave emission and tidal interaction. However, given our large mass uncertainties, the merger outcome of J0526$+$5934 is uncertain.

On the median and upper-end of our mass estimates, our data suggests that the most likely merger outcome of J0526$+$5934 is a ``dynamically driven double-degenerate double-detonation" (${\rm D}^6$) scenario in which unstable mass transfer ignites a Helium detonation near the surface of the accretor, which triggers a CO-core detonation and results in a sub-Chandrasekhar Type Ia supernova explosion of the accretor \citep{dan2012,dan2015,shen2018a,wong2023}. In this double-detonation scenario, the low-mass donor may survive its companion's explosion as a hyper-velocity star, retaining its orbital speed from before the explosion \citep[see][]{shen2018b,bauer2021,elbadry2023}. 

On the lower-end of our mass estimates, our data suggests that the merger of J0526$+$5934 is likely to result in a stable He-rich star \citep{zhang2014}, such as an R Coronae Borealis type star \citep{webbink1984}. This would naturally evolve into a massive CO white dwarf over time, contributing to the large fraction of merger products in the population of massive single white dwarfs \citep[see][]{cheng2020,kilic2023}.

\section{Summary \& Conclusions}\label{sec:conc}
In this work, we have presented our spectroscopic and photometric analysis of a new $P=20.506~{\rm min}$ ultra-compact LISA verification binary, independently discovered in the ZTF data archive and first reported in \citet{ren2023}.

We used archival Keck LRIS spectroscopy to estimate the atmospheric parameters of the visible component and find that, with $\log{g_2}=6.37\pm0.03$, the low-mass visible star is a post-core-burning hot subdwarf or an inflated low-mass He-core white dwarf. We performed light curve modeling to new multi-band high-speed photometry from the McDonald Observatory and find mass ratio $q=0.426^{+0.052}_{-0.050}$, mass $M_2=0.378^{+0.066}_{-0.060}~{\rm M_\odot}$, and volumetric radius $R_{2,{\rm vol}}=0.070\pm0.005~{\rm R_\odot}$, consistent with the estimates from our best-fitting SED model, $M_{2,{\rm SED}}=0.32^{+0.06}_{-0.05}~{\rm M_\odot}$ and $R_{2,{\rm SED}}=0.061^{+0.006}_{-0.005}~{\rm R_\odot}$.

We estimated the rate of orbital decay based on our most-probable system parameters and find that J0526$+$5934 will merge within $1.8\pm0.3~{\rm Myr}$ and most likely result in a ${\rm D^6}$ scenario supernova explosion or form a He-rich star that eventually evolves into a massive single white dwarf. While our mass estimates are uncertain, which results in large uncertainties in potential merger outcome, future timing measurements will provide a precise estimate to the chirp mass of J0526$+$5934 and help characterize the expected LISA gravitational wave signal, providing a clear solution to the eventual fate of J0526$+$5934.

\section*{Acknowledgements}

AK acknowledges support from NASA through grant 80NSSC22K0338. TK acknowledges support from the National Science Foundation through grant AST \#2107982, from NASA through grant 80NSSC22K0338 and from STScI through grant HST-GO-16659.002-A. We thank Andreas Irrgang for the development of the spectrum and SED-fitting tools and his contributions to the model atmosphere grids.

This work was supported in part by NSERC Canada and by the Fund FRQ-NT (Qu\'ebec).

The authors acknowledge the High Performance Computing Center at Texas Tech University for providing computational resources that have contributed to the research results reported within this paper.

We thank the anonymous referee for the comments and suggestions that greatly improved the quality of this work.

Based on observations obtained with the Samuel Oschin 48-inch Telescope at the Palomar Observatory as part of the Zwicky Transient Facility project. ZTF is supported by the National Science Foundation under grant No. AST-1440341 and a collaboration including Caltech, IPAC, the Weizmann Institute for Science, the Oskar Klein Center at Stockholm University, the University of Maryland, the University of Washington, Deutsches Elektronen-Synchrotron and Humboldt University, Los Alamos National Laboratories, the TANGO Consortium of Taiwan, the University of Wisconsin at Milwaukee, and Lawrence Berkeley National Laboratories. Operations are conducted by COO, IPAC, and UW.

This work has made use of data from the Asteroid Terrestrial-impact Last Alert System (ATLAS) project. The Asteroid Terrestrial-impact Last Alert System (ATLAS) project is primarily funded to search for near earth asteroids through NASA grants NN12AR55G, 80NSSC18K0284, and 80NSSC18K1575; byproducts of the NEO search include images and catalogs from the survey area. This work was partially funded by Kepler/K2 grant J1944/80NSSC19K0112 and HST GO-15889, and STFC grants ST/T000198/1 and ST/S006109/1. The ATLAS science products have been made possible through the contributions of the University of Hawaii Institute for Astronomy, the Queen’s University Belfast, the Space Telescope Science Institute, the South African Astronomical Observatory, and the Millennium Institute of Astrophysics (MAS), Chile.

This research has made use of the Keck Observatory Archive (KOA), which is operated by the W. M. Keck Observatory and the NASA Exoplanet Science Institute (NExScI), under contract with the National Aeronautics and Space Administration. 

Some of the data presented herein were obtained at the W. M. Keck Observatory, which is operated as a scientific partnership among the California Institute of Technology, the University of California, and the National Aeronautics and Space Administration. The Observatory was made possible by the generous financial support of the W. M. Keck Foundation.  

The authors wish to recognize and acknowledge the very significant cultural role and reverence that the summit of Maunakea has always had within the indigenous Hawaiian community.  We are most fortunate to have the opportunity to conduct observations from this mountain. 

The Pan-STARRS1 Surveys (PS1) and the PS1 public science archive have been made possible through contributions by the Institute for Astronomy, the University of Hawaii, the Pan-STARRS Project Office, the Max-Planck Society and its participating institutes, the Max Planck Institute for Astronomy, Heidelberg and the Max Planck Institute for Extraterrestrial Physics, Garching, the Johns Hopkins University, Durham University, the University of Edinburgh, the Queen's University Belfast, the Harvard-Smithsonian Center for Astrophysics, the Las Cumbres Observatory Global Telescope Network Incorporated, the National Central University of Taiwan, the Space Telescope Science Institute, the National Aeronautics and Space Administration under grant No. NNX08AR22G issued through the Planetary Science Division of the NASA Science Mission Directorate, the National Science Foundation grant No. AST-1238877, the University of Maryland, Eotvos Lorand University (ELTE), the Los Alamos National Laboratory, and the Gordon and Betty Moore Foundation.

This publication makes use of data products from the Wide-field Infrared Survey Explorer, which is a joint project of the University of California, Los Angeles, and the Jet Propulsion Laboratory/California Institute of Technology, funded by the National Aeronautics and Space Administration. 

\facilities{Struve (ProEM), Keck:I (LRIS)} 
\software{\textsc{astropy} \citep{astropy2013,astropy2018,astropy2022},
          \textsc{iraf} \citep{tody1986,tody1993}
          \textsc{ldasoft} \citep{littenberg2020},
          \textsc{lcurve} \citep{copperwheat2010},
          }
          
\bibliographystyle{aasjournal}



\end{document}